\documentclass[sigconf]{acmart}
\AtBeginDocument{%
  }

\usepackage{tabularx}
\usepackage{multirow}
\usepackage{colortbl}

\copyrightyear{2026}
\acmYear{2026}
\setcopyright{cc}
\setcctype{by}
\acmConference[MM '26] {Proceedings of the 34th ACM International Conference on Multimedia}{November 10--14, 2026}{Rio de Janeiro, Brazil.}
\acmBooktitle{Proceedings of the 34th ACM International Conference on Multimedia (MM '26), November 10--14, 2026, Rio de Janeiro, Brazil}
\acmISBN{979-8-4007-2213-4/2026/11}
\acmDOI{10.1145/3767308.3836048}
\begin{document}

\title{SPECTRA: On-Device Cognitive Perturbation and Trajectory Analysis for Autonomous Edge-Cloud GUI Grounding}

\author{Zhan Qu}
\email{quzhan@zju.edu.cn}
\orcid{0009-0009-5570-0884}
\affiliation{
  \institution{Zhejiang University}
  \city{Hangzhou}
  \country{China}
}

\author{Hui Zang}
\email{zanghui@huawei.com}
\orcid{0009-0003-5775-8648}
\affiliation{
  \institution{Huawei Technologies Co., Ltd.}
  \city{Shenzhen}
  \country{China}
}

\author{Ran Chen}
\email{chenran50@huawei.com}
\orcid{0000-0002-2770-4070}
\affiliation{
  \institution{Huawei Technologies Co., Ltd.}
  \city{Shenzhen}
  \country{China}
}

\author{Tao Wang}
\email{wangtao155@huawei.com}
\orcid{0009-0006-6319-3075}
\affiliation{
  \institution{Huawei Technologies Co., Ltd.}
  \city{Shenzhen}
  \country{China}
}

\author{Shengyu Zhang}
\correspondingauthor
\email{sy\_zhang@zju.edu.cn}
\orcid{0000-0002-0030-8289}
\affiliation{
  \institution{Zhejiang University}
  \city{Hangzhou}
  \country{China}
}

\renewcommand{\shortauthors}{Qu et al.}

\begin{abstract}
The effectiveness of edge-cloud collaboration for GUI grounding depends on autonomous requesting, where the edge agent selectively offloads complex tasks to the powerful cloud. However, in visually dense scenarios, lightweight edge agents often exhibit overconfident hallucinations, leading to a misalignment between confidence and accuracy that hinders reliable autonomous requesting. To address this, we leverage the observation that an agent's \textit{cognitive instability} leads to significant latent drift under minute perturbations due to steep decision boundaries. We propose SPECTRA, a lightweight autonomous request framework for edge-cloud GUI grounding, comprising (1) \textbf{\underline{S}}aliency-guided Targeted \textbf{\underline{P}}erturbation and (2) \textbf{\underline{E}}fficient \textbf{\underline{C}}ognitive \textbf{\underline{TR}}ajectory \textbf{\underline{A}}nalysis. SPECTRA conducts a visual cognitive stress test by injecting masks into critical visual anchors and quantifies the topological divergence of the agent's high-dimensional cognitive trajectories during the prefill phase, avoiding inefficient output decoding. Experiments demonstrate that SPECTRA performs cloud request assessment without autoregressive decoding. Our GTA1-32B+InfiGUI-G1-3B and GTA1-32B+Holo1.5-3B maintain 93.44\% and 95.60\% of cloud-only performance with average request rates of 37.58\% and 39.24\%, respectively.
\end{abstract}




\maketitle

\section{Introduction}
\begin{figure}[t]
    \centering
    \includegraphics[width=\linewidth]{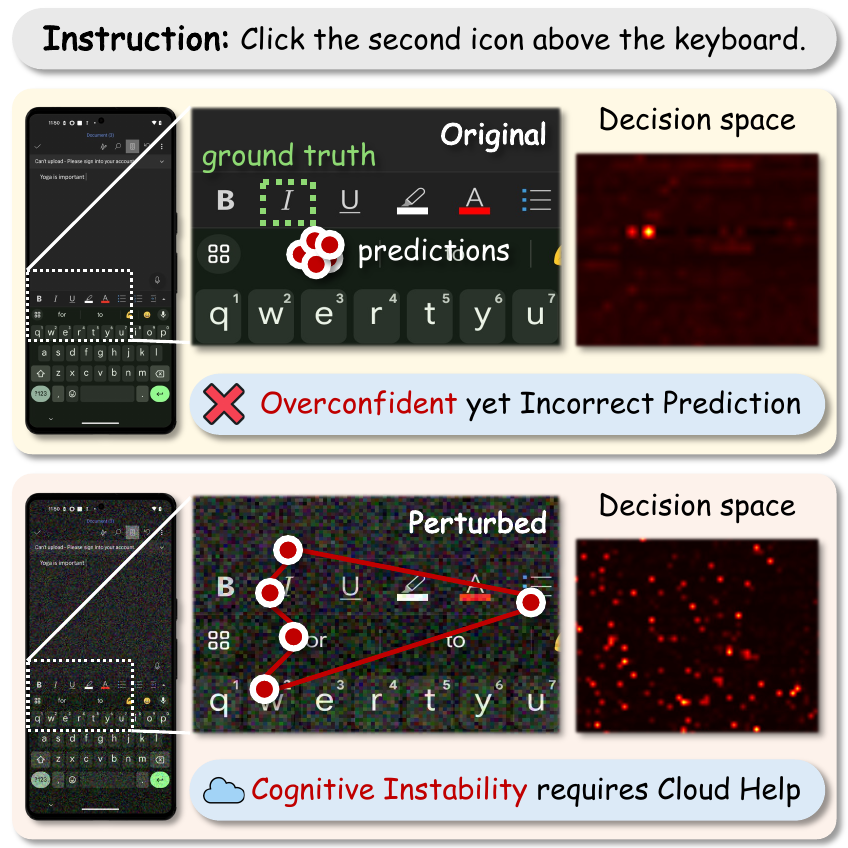}
    \caption{Cognitive instability in GUI grounding. Top: Under original input, the edge agent is overconfident. Although the predictions (red dots) miss the ground truth (green box), the decision space remains concentrated. Bottom: Under minute perturbations, the predictions exhibit spatial drift and decision space divergence. We leverage this instability to autonomously trigger cloud requests.}
    \Description{A two-part figure comparing the prediction results of a GUI agent on a smartphone keyboard interface for the instruction "click the second icon above the keyboard" under original and perturbed conditions. The top half is the original condition, showing the agent's predicted coordinates are a group of clustered red dots, deviating from the correct green dashed box, indicating the agent is overconfident but predicts incorrectly. The bottom half is the perturbed condition, showing the same interface after adding visual Gaussian noise. The red prediction dots are scattered across various icons and letters, indicating the agent's cognitive instability.}
    \label{fig:head}
\end{figure}

Multimodal large language model (MLLM)-based agents have demonstrated great potential in manipulating graphical user interfaces (GUIs)~\cite{rawles2025androidworld,xie2024osworld,chen2025pg,wu2025gui,niu2024screenagent}. A fundamental capability of these agents is GUI grounding~\cite{cheng2024seeclick}, which maps natural language instructions to the screen coordinates of target interactive elements, enabling agents to execute corresponding actions such as clicks. While cloud-based agents exhibit remarkable performance, users typically require real-time responses from agents to ensure a coherent experience in real-world GUI interaction scenarios. This growing demand for real-time interaction is accelerating the deployment of GUI agents to edge devices~\cite{wang2024mobile,vasu2025fastvlm,wu2024mobilevlm}.

However, edge agents with limited capabilities struggle to maintain grounding performance in visually dense GUI scenarios~\cite{li2025screenspot}. Therefore, edge-cloud collaboration~\cite{liu2025collaborative, zhao2025device, wang2025holotrace, wang2025multi} has become a promising solution, where autonomous requesting is required to automatically seek cloud assistance when a local task exceeds the agent's capabilities. Unlike natural images, GUIs consist of discrete and dense interactive elements, such as tightly packed tiny functional icons in professional software and structurally homogeneous list items in system interfaces. When faced with numerous irrelevant visual distractors, edge agents lack the fine-grained discriminative capability to lock onto the correct target. Instead, they erroneously collapse the probability distribution onto distractors that share partial visual features with the target, exhibiting overconfident hallucinations~\cite{li2025screenspot, mei2025can, zhang2025dhcp}. This misalignment between confidence and accuracy deprives the agent of the ability to identify its capability boundaries based on direct output statistics, posing a challenge to reliable autonomous requesting.

We address this challenge by leveraging empirical observations of the agent's \textbf{cognitive instability} in GUI grounding. We observe that while edge agents may output high-confidence coordinates based on priors, their anchoring on critical visual cues is often fragile. This fragility stems from the demand for precise visual matching in GUI grounding, requiring the edge agent to develop highly selective responses to specific visual details. Such sensitivity creates steep decision boundaries between the target and distractors. Essentially, an agent that truly comprehends the instruction can establish a robust alignment between the target and the global GUI context, whereas a guessing or hallucinating agent merely overfits to spurious visual correlations. Consequently, any latent cognitive instability is amplified, as illustrated in Figure~\ref{fig:head}: when visual cues are slightly perturbed, these fragile correlations collapse. This causes the agent to fail in maintaining prediction consistency, leading to significant spatial drift in the predicted coordinates. Although capturing the dispersion of output coordinates can trigger autonomous requests to the cloud, the reliance on multi-round decoding incurs computational costs unacceptable for edge devices.

Building on this insight, we propose an active introspection paradigm. Instead of passively observing the explicit output distribution, we actively stress test the agent's visual cognition. Specifically, we identify critical visual cues that the agent relies on during early reasoning and inject perturbations to challenge the robustness of its visual grounding. As recent studies have revealed~\cite{azaria2023internal}, an agent's internal activation patterns intrinsically encode an awareness of the truthfulness of its predictions even before any explicit outputs are generated. By injecting perturbations, we further amplify this inherent cognitive instability. Because this evaluation is conducted directly on hidden states, we can track the agent's cognitive trajectories without autoregressive decoding. Under perturbation, cognitive instability manifests as the dispersion of these trajectories in latent space, making topological divergence a measurable proxy for unreliable grounding. Greater divergence suggests that the agent is merely guessing or hallucinating. This allows us to identify potential failures early in the reasoning process before generating any specific coordinates. A request for cloud assistance can then be triggered, thereby avoiding costly decoding computation.

Technically, we implement this paradigm via SPECTRA, a lightweight autonomous request framework for edge-cloud GUI grounding. SPECTRA consists of two target designs: \textbf{(1) Saliency-Guided Targeted Perturbation.} It computes the norm saliency map during visual reasoning to identify critical visual grounding anchors, which are the specific GUI regions the agent relies on. Building on this, adversarial binary masks are injected into these regions to simulate stress tests on visual cognition. \textbf{(2) Efficient Cognitive Trajectory Analysis.} It performs parallel mask injections during the prefill stage and samples hidden states from the agent's language model layer. These hidden states form a set of high-dimensional cognitive trajectories. A lightweight trajectory predictor then analyzes the topological divergence patterns of these trajectories to quantify cognitive instability and drive autonomous requesting. Experiments demonstrate that SPECTRA performs cloud request assessment without autoregressive decoding, empowering the edge agent with introspection capabilities for reliable edge-cloud collaboration.

Our contributions are summarized as follows:
\begin{itemize}
    \item We propose SPECTRA, a lightweight autonomous request framework for edge-cloud GUI grounding. It establishes an active introspection paradigm for edge agents, identifying their capability boundaries by amplifying the inherent cognitive instability.
    \item We introduce Saliency-Guided Targeted Perturbation to simulate visual cognitive stress by masking salient GUI regions, and Efficient Cognitive Trajectory Analysis to quantify the topological divergence of the agent's cognitive trajectories without inefficient output decoding.
    \item Extensive evaluations demonstrate that SPECTRA performs reliable and efficient cloud request assessment. Our GTA1-32B+InfiGUI-G1-3B and GTA1-32B+Holo1.5-3B pairs maintain 93.44\% and 95.60\% of the cloud-only performance with average request rates of 37.58\% and 39.24\%, respectively.
\end{itemize}

\begin{figure*}[ht]
    \centering
    \includegraphics[width=\linewidth]{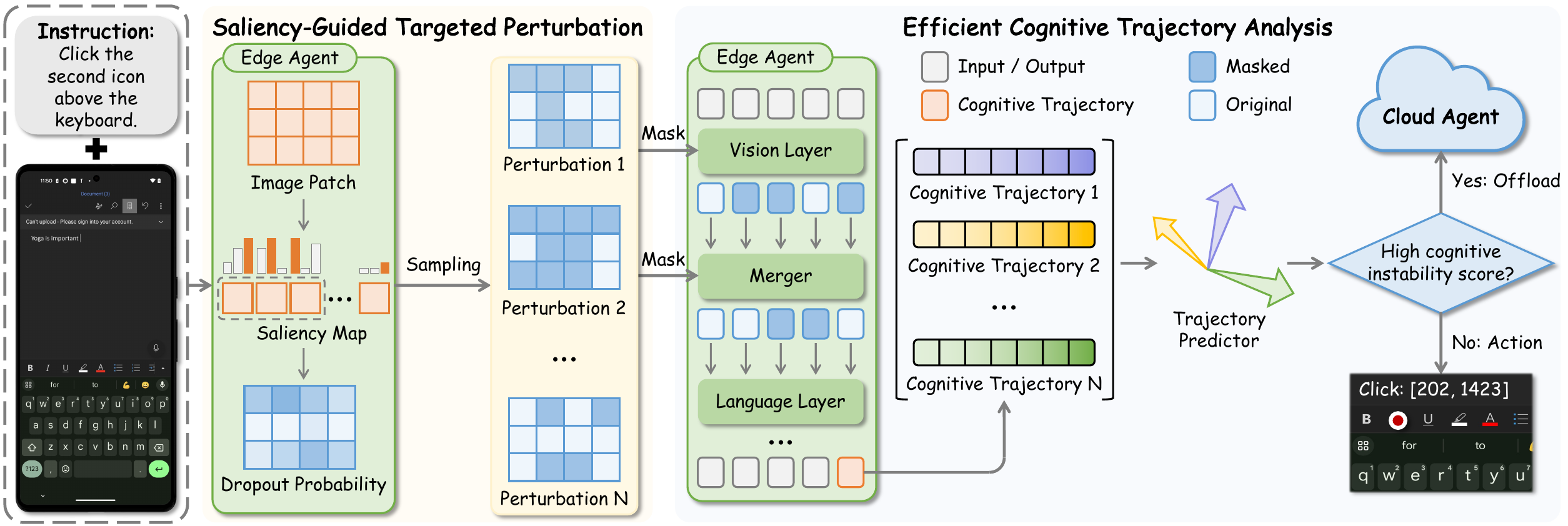}
    \caption{Overview of SPECTRA. Saliency-Guided Targeted Perturbation constructs adversarial masks on critical visual anchors to simulate cognitive stress. Subsequently, Efficient Cognitive Trajectory Analysis extracts hidden states as latent cognitive trajectories and quantifies their topological divergence to autonomously decide between local action and cloud requests.}
    \Description{The workflow of the SPECTRA. The instruction and smartphone GUI image are fed into the edge agent. Saliency-Guided Targeted Perturbation shows the vision layer processing image patches to create a saliency map, and generating multiple masks. Efficient Cognitive Trajectory Analysis applies the masks to the visual layer and merger, extracting N different cognitive trajectories. These trajectories are then fed into the trajectory predictor. Finally, it judges whether the cognitive instability score is higher than the threshold. If yes, the task is offloaded to the cloud agent. Otherwise, the edge agent executes locally.}
    \label{fig:framework}
\end{figure*}

\section{Related Work}
\subsection{GUI Agents and Grounding}
The emergence of MLLMs has changed the interaction paradigm of GUI agents. Early methods mainly operated GUIs using structured UI representations~\cite{lu2024omniparser,deng2023mind2web}, which are sometimes inaccessible or poorly maintained. Instead, MLLM-based agents can interact with GUIs directly through visual perception like humans~\cite{rawles2025androidworld,xie2024osworld,chen2025pg,wu2025gui,chen2025evaluating}. In this domain, GUI grounding introduced by SeeClick~\cite{cheng2024seeclick} aims to map natural language instructions to the precise spatial coordinates of target elements~\cite{liu2026infigui,hai2025holo15modelfamily,yang2025gta1}. To enhance grounding capability, researchers have explored supervised fine-tuning~\cite{gou2025navigating,qin2025ui} on specialized, large-scale GUI corpora~\cite{li2025screenspot,niu2024screenagent} for better visual-textual alignment, as well as reinforcement learning with rule-based rewards~\cite{liu2026infigui,yang2025gta1} to alleviate grounding ambiguity and insufficient exploration. Although large-scale cloud agents demonstrate remarkable performance, real-time GUI interaction has made lightweight on-device agents a mainstream trend~\cite{wang2024mobile,vasu2025fastvlm,wu2024mobilevlm,huang2025mvisu}. However, these edge agents often lack sufficient fine-grained discrimination in visually dense GUIs, leading to overconfident hallucinations. To address this limitation, edge-cloud collaboration offers a practical solution by enabling agents to autonomously seek cloud assistance when tasks exceed their capabilities.

\subsection{Edge-Cloud Collaboration}
Edge-cloud collaboration aims to approach cloud-level performance while maintaining edge-level latency through dynamic task allocation~\cite{lv2025collaboration,long2024diffusion,liu2025collaborative,zhao2025device,wang2025holotrace,wang2025multi}. It is widely applied in real-time and resource-constrained scenarios, such as GUI automation~\cite{yi2026ecoagent}, recommendation systems~\cite{lv2025collaboration,long2024diffusion} and video analytics~\cite{liu2025collaborative,kang2017noscope}. A core challenge is constructing an efficient autonomous request mechanism to determine when sufficiently complex tasks should be offloaded to the cloud~\cite{zhao2025device,wang2025holotrace}. Current LLM-based strategies usually rely on uncertainty estimation. Single-pass methods~\cite{malinin2021uncertainty,gupta2024language} typically calculate confidence or predictive entropy from output token probabilities but fail to reliably reflect prediction correctness under hallucinations in GUI grounding. Multi-sampling methods such as SelfCheckGPT~\cite{manakul2023selfcheckgpt} and EigenScore~\cite{chen2024inside} measure consistency across multiple generated responses. While more robust, repeated autoregressive decoding introduces significant overhead and unacceptable latency for real-time GUI operations. Probing internal model states for truthfulness without full decoding has recently shown promise~\cite{azaria2023internal,chen2024inside}, but existing approaches lack adaptation to visual-spatial GUI tasks. Consequently, GUI grounding still lacks a reliable and efficient collaboration mechanism that ensures accurate requesting and a seamless user experience.

\section{Method}
\subsection{Problem Formulation}
Formally, we define the GUI grounding task as follows: given a GUI screenshot $\mathcal{V} \in \mathbb{R}^{H \times W \times 3}$ and a natural language user instruction $\mathcal{T}$, the goal is to ground the target GUI element corresponding to $\mathcal{T}$. The agent $\mathcal{A}$ outputs a coordinate point $\hat{P} = (x, y)$ representing the predicted location on the screen. The grounding process can be formulated as:
\begin{equation}
    \hat{P} = \mathcal{A}(\mathcal{V}, \mathcal{T}).
\end{equation}

In edge-cloud collaboration, we consider two agents: a lightweight edge agent $\mathcal{A}_{e}$ with low latency but limited capacity, and a powerful cloud agent $\mathcal{A}_{c}$ that provides superior reasoning at higher costs. To balance efficiency and performance, the core objective is to design a decision function $\mathcal{G}$ to schedule collaboration. For each input $(\mathcal{V}, \mathcal{T})$, $\mathcal{G}$ outputs a binary decision $d \in \{0, 1\}$, where $d=1$ triggers a cloud request. The final prediction $\hat{P}^{*}$ is derived as:
\begin{equation}
    \hat{P}^{*} = (1-d) \cdot \mathcal{A}_{e}(\mathcal{V}, \mathcal{T}) + d \cdot \mathcal{A}_{c}(\mathcal{V}, \mathcal{T}).
\end{equation}

Our goal is to optimize $\mathcal{G}$ that minimizes the expected cloud request rate, subject to a reliability constraint $\delta$. Let $\mathcal{B}$ denote the ground truth bounding box corresponding to the input. The optimization problem can be formulated as:
\begin{equation}
    \label{equ:opt}
    \min_{\mathcal{G}} \mathbb{E}_{(\mathcal{V}, \mathcal{T}, \mathcal{B})} [d] \quad \text{s.t.} \quad \mathbb{E}_{(\mathcal{V}, \mathcal{T}, \mathcal{B})} \left[ \mathbb{I}(\hat{P}^{*} \in \mathcal{B}) \right] \ge \delta,
\end{equation}
where $\mathbb{I}(\cdot)$ is the indicator function that equals 1 if the condition holds. This formulation encapsulates the dual challenges of maximizing collaborative efficiency by minimizing cloud dependency, while ensuring grounding reliability.

\subsection{SPECTRA Design}
The design goal is to establish a lightweight autonomous request framework on the device that selectively offloads sufficiently difficult tasks to the cloud, reducing end-to-end latency while maintaining task performance. To this end, we propose SPECTRA, grounded in the insight that an unstable agent exhibits significant latent drift even under minute perturbations. As illustrated in Figure~\ref{fig:framework}, SPECTRA consists of two core components: (1) Saliency-Guided Targeted Perturbation, which conducts targeted cognitive stress tests based on the agent's visual understanding of the GUI; and (2) Efficient Cognitive Trajectory Analysis, which analyzes the topological divergence patterns of the agent's cognitive trajectories in stress tests to determine whether to offload to the cloud.

\subsubsection{Saliency-Guided Targeted Perturbation}
Unlike natural images with continuously distributed visual features, GUI screenshots are composed of discrete and dense interactive elements scattered amidst background noise. Although these elements are dispersed in pixel space, they often cluster in the high-dimensional manifold, leading to partial feature sharing between the target and similar distractors. This characteristic requires the agent to perform highly selective grounding during reasoning, resulting in steep decision boundaries in the decision space. In this process, the agent activates specific GUI regions for prediction, defined as visual anchors. Essentially, a cognitively stable agent can maintain focus on these anchors, preserving cross-modal alignment under external perturbations. Based on this agent cognitive characteristic determined by GUI topology, we introduce Saliency-Guided Targeted Perturbation, which constructs adversarial masks to stress test the agent.

We leverage the intrinsic activation patterns of the edge agent $\mathcal{A}_{e}$ to instantiate visual anchors. Let $\mathbf{H} = \{h_1, h_2, \dots, h_L\} \in \mathbb{R}^{L \times D}$ denote the hidden states in visual reasoning, where $L$ is the sequence length and $D$ is the hidden dimension. We compute a saliency map based on the saliency score $s_i$ of each image patch $h_i$, defined as its $L_2$ norm. This map serves as a quantitative proxy for the agent's visual focus on the GUI screenshot:
\begin{equation}
    s_i = \|h_i\|_2,
\end{equation}
where $\|\cdot\|_2$ denotes the $L_2$ norm. Patches with high $s_i$ values are identified as visual anchors, representing the critical visual cues for the current grounding task.

To ensure spatial continuity in perturbation, we introduce a sliding window mechanism. If a specific image patch corresponds to a highly salient region, its neighbors are likely also very important. Therefore, we assign the maximum saliency score within the window to the central patch. We implement this operation directly along the flattened token sequence, given that vertical contexts are already embedded during visual encoding. This mechanism achieves approximate local smoothness without the computational overhead of reconstructing 2D spatial grids. The contextualized saliency score $s_i^\prime$ is calculated as:
\begin{equation}
    s_i^\prime = \max\{ s_j \mid j \in [\max(1, i-w), \min(L, i+w)] \},
\end{equation}
where $w$ denotes the radius of the sliding window.

Based on this, we map the contextualized scores to masking probabilities to determine the perturbation intensity. The mask probability $p_i$ for each image patch is derived via the following scaled min-max normalization:
\begin{equation}
    p_i = \lambda \cdot \frac{s_i^\prime - s_{\text{min}}^\prime}{s_{\text{max}}^\prime - s_{\text{min}}^\prime + \epsilon},
\end{equation}
where $\lambda \in [0, 1]$ denotes the perturbation intensity coefficient, $\epsilon$ is a small constant for numerical stability, and $s_{\text{min}}^\prime = \min(\mathbf{s}^\prime)$ and $s_{\text{max}}^\prime = \max(\mathbf{s}^\prime)$ represent the minimum and maximum scores within the saliency sequence $\mathbf{s}^\prime = \{s_1^\prime, \dots, s_L^\prime\}$, respectively. 

We use $\mathbf{p} = \{p_1, \dots, p_L\}$ to sample binary masks. For the $k$-th parallel stress test, the binary mask $\mathbf{M}^k = \{\mathbf{m}_1^k, \dots, \mathbf{m}_L^k\} \in \{0, 1\}^{L \times D}$ is sampled independently across feature dimensions according to the corresponding patch probability:
\begin{equation}
    P(m_{i,d}^k = 1) = p_i, \quad d=1,\dots,D.
\end{equation}
Once sampled, the mask is applied to the original hidden states to simulate localized cognitive bottlenecks. The perturbed hidden state $\tilde{h}_i^k$ is computed using saliency-conditioned feature dropout:
\begin{equation}
    \tilde{h}_i^k = \frac{h_i \odot (\mathbf{1} - \mathbf{m}_i^k)}{1 - p_i + \epsilon},
\end{equation}
where $\odot$ denotes element-wise multiplication. We generate $N$ independent masks $\{\mathbf{M}^1, \dots, \mathbf{M}^N\}$, where each mask serves as an independent stress test, challenging the agent to maintain robust reasoning under these adversarial conditions.

\begin{table*}[t]
    \centering
    \caption{Quantitative comparison of the performance of request decision and collaboration across three benchmarks. The best and second-best results are highlighted in bold and underlined, respectively.}
    \label{tab:main_results}
    \begin{tabularx}{\textwidth}{l|*{3}{>{\centering\arraybackslash}X}|*{3}{>{\centering\arraybackslash}X}|*{3}{>{\centering\arraybackslash}X}}
        \toprule
        \multirow{2}{*}{\textbf{Methods}} & \multicolumn{3}{c|}{\textbf{MMBench-GUI}} & \multicolumn{3}{c|}{\textbf{ScreenSpot-Pro}} & \multicolumn{3}{c}{\textbf{UI-I2E-Bench}} \\
        & \textbf{AUC} $\uparrow$ & \textbf{SRCC} $\uparrow$ & \textbf{AUCG} $\uparrow$ & \textbf{AUC} $\uparrow$ & \textbf{SRCC} $\uparrow$ & \textbf{AUCG} $\uparrow$ & \textbf{AUC} $\uparrow$ & \textbf{SRCC} $\uparrow$ & \textbf{AUCG} $\uparrow$ \\
        \midrule
        
        \multicolumn{10}{c}{\textit{Model: InfiGUI-G1-3B}} \\
        \midrule
        Random & 51.05 & 1.86 & 52.30 & 49.90 & 1.62 & 48.69 & 47.69 & -4.26 & 47.22 \\
        LN-Confidence & 57.37 & 9.49 & 53.03 & 62.82 & 8.69 & 53.27 & 51.65 & -1.66 & 38.24 \\
        LN-Entropy & \underline{63.11} & \underline{17.82} & 57.21 & 65.08 & \underline{16.77} & 53.62 & 54.60 & 4.03 & 48.84 \\
        CoT & 60.56 & 17.81 & 57.41 & 56.89 & 14.49 & 49.45 & \underline{58.22} & \underline{14.05} & \underline{69.58} \\
        EigenScore & 55.77 & 8.38 & 56.02 & 56.86 & 10.08 & 52.15 & 53.90 & 5.95 & 53.29 \\
        Self-Consistency & 61.30 & 15.55 & \underline{63.44} & \underline{66.65} & 15.16 & \textbf{57.03} & 57.44 & 7.50 & 45.57 \\
        
        \rowcolor{gray!10}
        \textbf{SPECTRA (Ours)} & \textbf{74.57} & \textbf{35.44} & \textbf{68.40} & \textbf{69.15} & \textbf{36.72} & \underline{56.32} & \textbf{67.29} & \textbf{22.60} & \textbf{71.39} \\
        
        \midrule
        
        \multicolumn{10}{c}{\textit{Model: Holo1.5-3B}} \\
        \midrule
        Random & 50.12 & -0.74 & 52.25 & 49.96 & -1.84 & 52.45 & 49.55 & -2.69 & 51.15 \\
        LN-Confidence & 49.22 & -9.62 & 54.60 & 70.82 & 5.72 & 72.18 & 48.75 & -10.34 & 40.65 \\
        LN-Entropy & \underline{69.26} & \underline{19.05} & \underline{69.43} & \textbf{83.62} & \textbf{26.58} & \textbf{79.79} & \underline{69.13} & \underline{21.09} & \underline{60.55} \\
        CoT & - & - & - & - & - & - & - & - & - \\
        EigenScore & 62.28 & 12.94 & 63.41 & 72.88 & 16.48 & 69.37 & 58.22 & 10.38 & 50.49 \\
        Self-Consistency & 61.18 & 10.41 & 63.47 & 73.90 & 21.24 & \underline{73.28} & 56.70 & 5.39 & 48.26 \\
        
        \rowcolor{gray!10}
        \textbf{SPECTRA (Ours)} & \textbf{75.62} & \textbf{34.92} & \textbf{70.25} & \underline{76.54} & \underline{26.05} & 65.92 & \textbf{72.37} & \textbf{32.52} & \textbf{63.82} \\
        
        \bottomrule
    \end{tabularx}
\end{table*}

\subsubsection{Efficient Cognitive Trajectory Analysis}
Existing collaborative decision methods typically rely on calculating uncertainty statistics on generated token sequences or final output results~\cite{yue2024large,gupta2024language}. While these methods are effective, autoregressive decoding leads to prohibitive overhead, failing to meet the real-time interaction demands of GUI scenarios. To overcome this bottleneck, we shift the assessment perspective from the explicit output space to the implicit latent space. Inspired by recent research finding that internal activation patterns can encode model truthfulness~\cite{azaria2023internal,chen2024inside}, we posit that the agent's confidence can be intrinsically captured from its hidden states. Building on this, we propose Efficient Cognitive Trajectory Analysis, which triggers autonomous requests to the cloud by analyzing the topological patterns of the agent's cognitive trajectories under multiple perturbations without decoding.

Specifically, we perform a batch of $N$ parallel forward passes during the prefill phase. In each pass, different sampled masks are applied to the hidden state outputs of specific layers of the visual encoder and the cross-modal merger to simulate diverse visual cognitive stresses. From these parallel passes, we then extract the hidden states of the last token from the penultimate layer of the language model backbone as proxies for the agent's cognitive trajectories. The penultimate layer retains rich semantic and cross-modal alignment information. Earlier layers lack sufficient reasoning depth, while the final layer is specialized for token decoding over the vocabulary space. Furthermore, the last token aggregates a comprehensive analysis of the multimodal prompt~\cite{ren2023out,azaria2023internal}. This process yields a set of cognitive trajectory vectors $\mathcal{H} = \{z_1, z_2, \dots, z_N\}$, where $z_k \in \mathbb{R}^D$ represents the agent's cognitive trajectory under the $k$-th perturbation. Due to the high parallelism of the prefill phase, this strategy incurs substantially lower latency than complete decoding.

Geometrically, if the agent is guessing or hallucinating, minute perturbations will cause these hidden states to drift across the latent manifold, manifesting as divergent cognitive trajectories. To quantify this topological divergence, we employ a lightweight trajectory predictor $f_{\theta}(\cdot)$. To precisely capture the relational patterns among these perturbed cognitive trajectories, $f_{\theta}(\cdot)$ is parameterized as a transformer encoder. Each trajectory is first mapped through a learned projection as $\bar{z}_k=g(z_k)$, after which a learnable classification token $z_{\text{cls}}$ is prepended to form the input sequence $\mathbf{Z}^{(0)}=[z_{\text{cls}},\bar{z}_1,\dots,\bar{z}_N]$. Since the parallel perturbations are independent and unordered, we omit positional encodings to maintain permutation invariance.

The sequence is processed through $K$ layers of multi-head self-attention and multilayer perceptrons, which are denoted as $\text{MSA}$ and $\text{MLP}$ respectively. For the $l$-th layer, the forward propagation is formulated as:
\begin{align}
  \mathbf{\tilde{Z}}^{(l)} &= \text{LN}(\text{MSA}(\mathbf{Z}^{(l-1)}) + \mathbf{Z}^{(l-1)}), \\
  \mathbf{Z}^{(l)} &= \text{LN}(\text{MLP}(\mathbf{\tilde{Z}}^{(l)}) + \mathbf{\tilde{Z}}^{(l)}),
\end{align}
where $\text{LN}(\cdot)$ denotes layer normalization. The multi-head self-attention allows the model to compute the pairwise cross-correlation between different perturbed trajectories, thereby quantifying the latent dispersion, which indicates whether the trajectories tightly cluster or drift randomly. Finally, the state of the classification token from the last layer, $z_{\text{cls}}^{(K)}$, is extracted and passed through a linear projection head with a sigmoid activation $\sigma(\cdot)$ to output the estimated probability of a necessary cloud request, denoted as $\hat{y}$:
\begin{equation}
    \hat{y} = f_{\theta}(\mathcal{H}) = \sigma(\mathbf{W} z_{\text{cls}}^{(K)} + b),
\end{equation}
where $\mathbf{W}$ and $b$ are learnable weights and biases, respectively.

\paragraph{\normalfont\textbf{Predictor Optimization.}}
To enable the trajectory predictor $f_{\theta}(\cdot)$ to assess the edge agent's competence, we employ a supervised training strategy using labels derived from the agent's actual performance. For each training sample $(\mathcal{V}, \mathcal{T})$ with a ground truth bounding box $\mathcal{B}$, the edge agent performs clean inference to generate predicted coordinates $\hat{P}_e = \mathcal{A}_e(\mathcal{V}, \mathcal{T})$. The label $y^*$ is defined based on the correctness of this prediction:
\begin{equation}
    y^* = \begin{cases} 
    0, & \text{if } \hat{P}_e \in \mathcal{B} \quad (\text{Correct}), \\
    1, & \text{if } \hat{P}_e \notin \mathcal{B} \quad (\text{Incorrect}),
    \end{cases}
\end{equation}
where $y^*=1$ indicates a necessary cloud request. This process automatically generates a dataset of $\{(\mathcal{H}, y^*)\}$ pairs.

The predictor $f_{\theta}(\cdot)$ is designed to estimate the probability of a cloud request. During the training phase, we optimize the predictor's parameters $\theta$ by minimizing the binary cross-entropy loss:
\begin{equation}
    \mathcal{L}(\theta) = - \mathbb{E}_{(\mathcal{H}, y^*)} \left[ y^* \log \hat{y} + (1 - y^*) \log (1 - \hat{y}) \right].
\end{equation}

\paragraph{\normalfont\textbf{Requesting Decision.}}
During deployment, the trained predictor outputs the cognitive instability score $\rho = f_{\theta}(\mathcal{H})$ in real-time based on the extracted cognitive trajectories. The final binary decision $d$ is determined by comparing this score with a decision threshold $\tau$:
\begin{equation}
d = \mathbb{I}(\rho > \tau).
\end{equation}
If $d=1$, the task is offloaded to the cloud agent $\mathcal{A}_{c}$. Otherwise, the local prediction from the edge agent $\mathcal{A}_{e}$ is adopted. The threshold $\tau$ is calibrated on a held-out validation set by maximizing Youden's J statistic~\cite{fluss2005estimation}, ensuring an optimal balance between reliability and efficiency. Furthermore, in practical deployment, the threshold supports dynamic adaptation. For instance, $\tau$ can be increased to limit offload traffic when the cloud server is under high load.

\begin{figure*}[t]
    \centering
    \includegraphics[width=\linewidth]{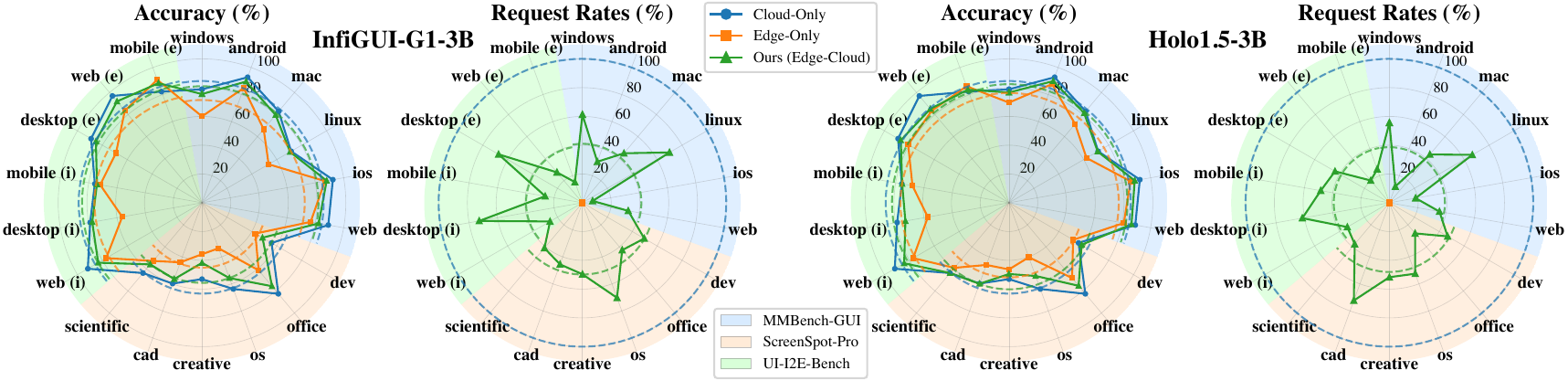}
    \caption{Accuracy and request rates of cloud-only, edge-only, and our method (at optimal thresholds) across domains on three benchmarks. Note: On the right, the edge-only baseline appears as a single point due to its zero request rate. Dashed arcs indicate benchmark averages.}
    \Description{Four radar charts compare the performance of InfiGUI-G1-3B and Holo1.5-3B across different domains in MMBench-GUI, ScreenSpot-Pro, and UI-I2E-Bench. For each model, the first chart shows the accuracy, where Ours (Edge-Cloud) is close to the upper bound of Cloud-Only and outperforms Edge-Only. The second chart shows the request rates, where our method has different request rates under different domains.}
    \label{fig:radar_charts_comparison}
\end{figure*}

\section{Experiments}
\subsection{Experimental Settings}
\subsubsection{Datasets.}
We conducted evaluations on three public GUI grounding benchmarks:
(1) \textbf{MMBench-GUI}~\cite{wang2026mmbench}: A benchmark for cross-platform interaction, spanning six major operating systems. It contains instructions of varying complexity and heterogeneous UI layouts, validating the generalizability across diverse GUI styles.
(2) \textbf{ScreenSpot-Pro}~\cite{li2025screenspot}: A benchmark for fine-grained GUI element grounding, containing high-resolution screenshots from multiple desktop platforms. This dataset demands precise grounding of target elements among dense distractors.
(3) \textbf{UI-I2E-Bench}~\cite{liu2025ui}: A benchmark focusing on instruction semantic understanding, containing a higher proportion of implicit user instructions. It specifically challenges the robustness of cross-modal reasoning and target alignment when facing non-intuitive descriptions.

\subsubsection{Evaluation Metrics.}
We employ two sets of metrics to assess the performance of request decision and collaboration, respectively.
(1) Request Decision Metrics. We use the widely adopted Area Under the Receiver Operating Characteristic Curve (\textbf{AUC}) to measure the capability in distinguishing between necessary cloud requests and edge-solvable tasks based on the predicted scores. We further employ Spearman's Rank Correlation Coefficient (\textbf{SRCC}) to assess the monotonic correlation between these scores and the spatial error distance of predicted coordinates from the ground truth.
(2) Collaborative Performance Metric. We propose the Area Under the Collaborative Gain (\textbf{AUCG}), which calculates the integral of the Normalized Collaborative Gain (NCG) over the request rate range $r \in [0, 1]$. NCG quantifies the proportion of the performance gap closed by agent collaboration relative to the cloud-only upper bound, which is given by:
\begin{equation}
    \text{NCG}(r) = \frac{\text{Acc}_{\text{col}}(r) - \text{Acc}_{\text{e}}}{\text{Acc}_{\text{c}} - \text{Acc}_{\text{e}}},
\end{equation}
where $\text{Acc}_{\text{e}}$ and $\text{Acc}_{\text{c}}$ denote the accuracy of the edge and cloud agents, respectively, and $\text{Acc}_{\text{col}}(r)$ represents the collaborative accuracy at a request rate $r$. A higher AUCG indicates superior accuracy retention at lower cloud costs.

\subsubsection{Compared Baselines.}
We compare our framework with representative baselines categorized into three groups: (1) \textbf{Random}, which assigns random scores to serve as a performance lower bound. (2) Single-Pass methods, including Length-Normalized Confidence (\textbf{LN-Confidence}) and \textbf{LN-Entropy}~\cite{malinin2021uncertainty} derived from output token log probabilities, and \textbf{CoT}~\cite{kadavath2022language}, which prompts the agent to output a confidence score alongside its predicted coordinates for self-evaluation. (3) Multi-Sampling methods, specifically \textbf{EigenScore}~\cite{chen2024inside} and \textbf{Self-Consistency}~\cite{wang2023self}. We adapt Self-Consistency to measure prediction instability by generating multiple coordinate predictions with temperature-scaled sampling at $t=0.4$ and using their spatial standard deviation as the uncertainty score.

\begin{figure}[t]
    \centering
    \includegraphics[width=\linewidth]{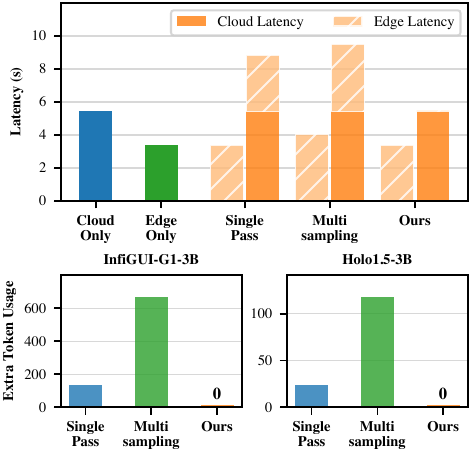}
    \caption{Cost analysis of latency and token usage. Top: Latency breakdown for local and cloud execution paths. Bottom: Extra token usage for request decision-making.}
    \Description{Bar charts comparing latency and extra token usage. The top chart shows our method adds lower edge latency compared to single-pass and multi-sampling baselines. The bottom charts show our method requires 0 extra tokens for decision-making, whereas baselines require up to hundreds.}
    \label{fig:cost_analysis}
\end{figure}

\subsubsection{Implementation Details.}
We employ InfiGUI-G1-3B~\cite{liu2026infigui} and Holo1.5-3B~\cite{hai2025holo15modelfamily} as the edge agents $\mathcal{A}_{e}$, and GTA1-32B~\cite{yang2025gta1} as the cloud agent $\mathcal{A}_{c}$. We apply perturbations with $\lambda=0.1$ to layers 7, 15, and 23 of the visual encoder, as well as to the cross-modal merger. The sliding window is applied during training with $w=2$. For both our method and the multi-sampling baselines, we set the number of samples to $N=5$. The trajectory predictor $f_{\theta}(\cdot)$ is based on a transformer encoder.

We construct a hybrid dataset containing approximately 7.3k samples, primarily sampled from subsets of Widget Caption~\cite{li2020widget} and OmniAct~\cite{kapoor2024omniact}. This dataset is split into 70\%/15\%/15\% for training, validation, and testing. We augment the training set with 100 few-shot samples from each evaluation benchmark using a 10$\times$ oversampling strategy. These samples are excluded from the final evaluation. We train $f_{\theta}(\cdot)$ for 100 epochs using AdamW optimizer~\cite{loshchilov2019decoupled} with a learning rate of $1e^{-5}$ and a batch size of 32.

\begin{figure}[t]
    \centering
    \includegraphics[width=\linewidth]{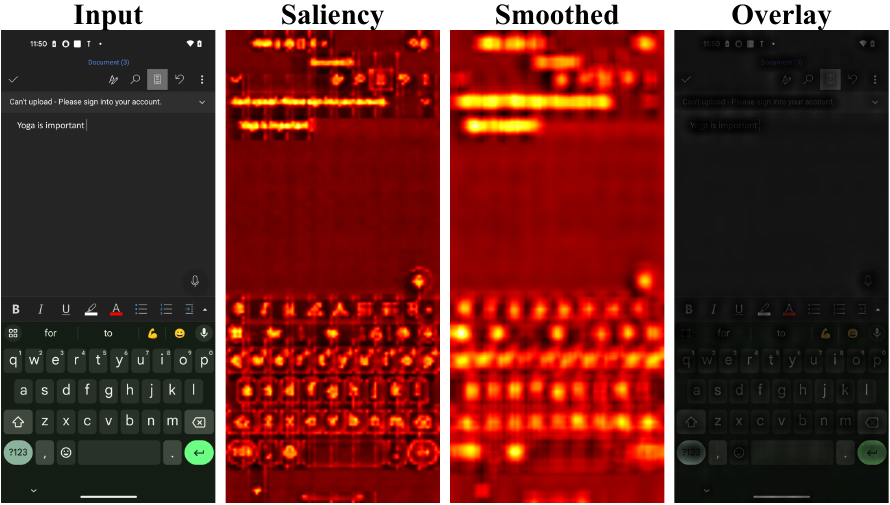}
    \caption{Saliency-Guided Targeted Perturbation on a document editing task. From left to right: original input, raw saliency map, smoothed saliency map, and final overlay.}
    \Description{A sequence of four images showing the target perturbation process on a mobile document editor. From left to right: the original input, the raw saliency map, the smoothed saliency map, and the final overlay.}
    \label{fig:saliency_vis}
\end{figure}

\subsection{Main Results and Analysis}
\subsubsection{Request Decision Performance.} 
Table~\ref{tab:main_results} compares request decision performance across three benchmarks and two edge agents. Our method consistently outperforms competing approaches across most metrics. Specifically, on MMBench-GUI, our framework surpasses the best baseline on InfiGUI-G1-3B by 11.46\% in AUC and 17.62\% in SRCC. It is interesting that LN-Entropy shows strength on ScreenSpot-Pro with Holo1.5-3B. The high element density in ScreenSpot-Pro confronts agents with multiple decision choices, making the entropy of the output probability distribution effective. However, LN-Entropy exhibits instability across diverse tasks, evidenced by significant performance drops on the semantically complex UI-I2E-Bench. This suggests that token-level probability distributions are often decoupled from cross-modal understanding, making them unreliable when instructions demand deep semantic reasoning rather than visual pattern matching. Furthermore, although Self-Consistency achieves relatively stable performance, its passive sampling is insufficient to probe the agent's cognitive instability, often leaving overconfident hallucinations undetected. In contrast, our method effectively identifies necessary cloud requests, validating the robustness of its decisions for autonomous requesting by edge agents.

\subsubsection{Collaborative Performance.}
Beyond request decision performance, we further evaluate the holistic collaborative benefit. As shown in Table~\ref{tab:main_results}, our method achieves the highest AUCG scores on both MMBench-GUI and UI-I2E-Bench. This achievement primarily stems from the superior request decision capability of our method. While some baselines achieve competitive AUCG on ScreenSpot-Pro, collaborative gain depends on both the request decision and the cloud agent's accuracy on the specifically offloaded tasks. For ScreenSpot-Pro, the tasks offloaded by baselines are effectively solved by the cloud agent. However, their collaborative performance lacks robustness across diverse domains. For instance, on UI-I2E-Bench with Holo1.5-3B, LN-Entropy drops to 60.55\% and Self-Consistency to 48.26\%. By comparison, SPECTRA maintains consistently high efficacy across all benchmarks. This demonstrates that our active introspection paradigm provides a robust foundation for reliable edge-cloud collaboration.

\subsubsection{Performance across Domains.}
Figure~\ref{fig:radar_charts_comparison} visualizes performance across domains. The edge-cloud collaboration demonstrates remarkable performance: the GTA1-32B+InfiGUI-G1-3B and GTA1-32B+Holo1.5-3B pairs maintain 93.44\% and 95.60\% of the cloud-only performance, with average request rates of only 37.58\% and 39.24\%, respectively. Crucially, the request distribution aligns with the edge agent's capability boundaries. In simple GUI scenarios like \textit{iOS} on MMBench-GUI, where InfiGUI-G1-3B achieves 86.80\% accuracy, our method suppresses cloud requests to 7.19\%. In challenging tasks such as the \textit{OS} subset of ScreenSpot-Pro, where the edge agent's accuracy drops to 33.67\%, our method raises the cloud request rate to 70.27\% to ensure reliability. These results validate that SPECTRA effectively aligns request decisions with edge capabilities, ensuring cloud resources are utilized only when necessary.

\begin{figure}[t]
    \centering
    \includegraphics[width=\linewidth]{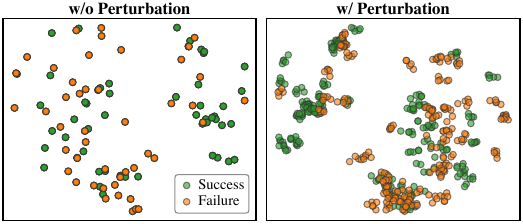}
    \caption{t-SNE visualization of cognitive trajectories. The left and right show the latent distributions without and with perturbation, respectively.}
    \Description{Two t-SNE scatter plots comparing successful and failed samples. The left plot (without perturbation) shows blurred boundaries between the two groups. The right plot (with perturbation) shows failed samples becoming more dispersed and visually distinct from the successful ones.}
    \label{fig:tsne_visualization}
\end{figure}

\subsubsection{Cost Analysis.}
Figure~\ref{fig:cost_analysis} shows a detailed comparison of latency and extra token usage for cloud request assessment between our design and different baselines. Benefiting from parallel prefill, our method maintains latency comparable to baselines in the local execution path. Notably, in the cloud execution path, baseline methods necessitate time-consuming local decoding. By comparison, completing cloud request assessment during the prefill phase allows our method to reduce edge latency by over 97\% to a negligible level. Regarding extra token usage, baseline methods require generating full response sequences before request decision. For instance, on InfiGUI-G1-3B, the multi-sampling baseline requires generating over 660 extra tokens. Our method achieves zero extra token decoding costs. This lightweight characteristic makes SPECTRA highly suitable for edge device applications.

\begin{table}[t]
    \centering
    \caption{Real on-device edge cost on ScreenSpot-Pro with InfiGUI-G1-3B. p95 denotes 95th-percentile latency.}
    \label{tab:ondevice_cost}
    \begin{tabularx}{\linewidth}{l>{\centering\arraybackslash}Xc>{\centering\arraybackslash}X}
        \toprule
        \textbf{$N$} & \textbf{Latency (s)} & \textbf{p95 (s)} & \textbf{Compute (TFLOPs)} \\
        \midrule
        1 & $4.71 \pm 0.35$ & 5.26 & $14.99 \pm 0.37$ \\
        3 & $5.50 \pm 0.37$ & 6.19 & $44.98 \pm 1.12$ \\
        5 & $6.44 \pm 0.46$ & 7.07 & $74.97 \pm 1.86$ \\
        7 & $6.79 \pm 0.37$ & 7.39 & $104.95 \pm 2.60$ \\
        \bottomrule
    \end{tabularx}
\end{table}

Table~\ref{tab:ondevice_cost} reports real on-device edge costs under different numbers of perturbations. As $N$ increases, compute grows linearly while mean latency increases sublinearly. Benefiting from parallel prefill, the mean latency at $N=5$ increases by only 36.7\% relative to $N=1$. The complete edge deployment occupies approximately 7.97 GB of VRAM. We further evaluate the practical overhead of the trajectory predictor, which adds only approximately 1.9 ms of latency and 46 MB of VRAM, making its overhead negligible.

\subsection{Visualization Results}
\subsubsection{Targeted Perturbation.} 
Figure~\ref{fig:saliency_vis} shows the intermediate steps of Saliency-Guided Targeted Perturbation on a document editing interface. High-norm regions are concentrated in element-dense areas such as the top toolbar and bottom keyboard, while the central blank document background has almost no activation. The sliding window further smooths the discrete high-activation points, and the overlay shows that the generated masks cover key interactive elements rather than irrelevant backgrounds. By avoiding uniform random masking, this targeted approach preserves the global GUI layout while challenging the fine-grained visual features the agent relies on for precise coordinate prediction.

\subsubsection{Cognitive Trajectories.} 
Figure~\ref{fig:tsne_visualization} presents the t-SNE visualization~\cite{van2008visualizing} of latent cognitive trajectories for 50 randomly sampled pairs of successful and failed samples. In the unperturbed state, successful and failed samples exhibit some distribution differences, but the boundary remains blurred, making reliable request decisions directly from static hidden states difficult. After perturbation, failed samples exhibit greater dispersion than successful ones: the relatively clustered trajectories of correct predictions reflect stable cross-modal alignment, while the scattered trajectories of failed samples reveal a lack of robust feature anchoring. These observations indicate that our trajectory predictor learns from dual cues: the potential semantic separation between samples and, more critically, the amplification of topological divergence caused by perturbation.

\begin{table}[t]
    \centering
    \caption{Ablation study on MMBench-GUI.}
    \label{tab:ablation}
    \begin{tabularx}{\linewidth}{l>{\centering\arraybackslash}X>{\centering\arraybackslash}X>{\centering\arraybackslash}X}
        \toprule
        \textbf{Methods} & \textbf{AUC} & \textbf{SRCC} & \textbf{AUCG} \\
        \midrule
        w/o Perturbation & 71.09 & 29.27 & 65.78 \\
        w/o Sliding Window & 72.51 & 31.18 & 66.07 \\
        Gaussian Noise & 71.01 & 32.21 & 67.97 \\
        Random Masking & 69.71 & 29.57 & 63.23 \\
        Inverse Saliency & 71.81 & 33.70 & 67.30 \\
        \midrule
        w/o Trajectory Predictor & 58.46 & 11.88 & 58.30 \\
        Mean + MLP & 69.58 & 30.25 & 65.12 \\
        Concatenation + MLP & 67.44 & 27.54 & 64.18 \\
        \midrule
        \rowcolor{gray!10}
        \textbf{Ours} & \textbf{74.57} & \textbf{35.44} & \textbf{68.40} \\
        \bottomrule
    \end{tabularx}
\end{table}

\begin{table}[t]
    \centering
    \caption{Trajectory extraction ablation on MMBench-GUI.}
    \label{tab:extraction_strategy}
    \begin{tabularx}{\linewidth}{l>{\centering\arraybackslash}X>{\centering\arraybackslash}X>{\centering\arraybackslash}X}
        \toprule
        \textbf{Extraction Strategy} & \textbf{AUC} & \textbf{SRCC} & \textbf{AUCG} \\
        \midrule
        \multicolumn{4}{c}{\textit{Extraction Layers (with Last Token)}} \\
        \midrule
        First Layer & 58.21 & 12.19 & 57.94 \\
        Middle Layer & 66.85 & 25.43 & 63.57 \\
        Last Layer & 73.81 & 34.12 & 67.94 \\
        \rowcolor{gray!10}
        \textbf{Penultimate Layer (Ours)} & \textbf{74.57} & \textbf{35.44} & \textbf{68.40} \\
        \midrule
        \multicolumn{4}{c}{\textit{Token Aggregation (on Penultimate Layer)}} \\
        \midrule
        Average Pooling & 71.16 & 29.85 & 65.92 \\
        \rowcolor{gray!10}
        \textbf{Last Token (Ours)} & \textbf{74.57} & \textbf{35.44} & \textbf{68.40} \\
        \bottomrule
    \end{tabularx}
\end{table}

\subsection{Ablation Study}
We conduct ablation studies on MMBench-GUI with InfiGUI-G1-3B, as summarized in Table~\ref{tab:ablation} and Table~\ref{tab:extraction_strategy}.

\subsubsection{Impact of Perturbation Strategy.}
The \textit{w/o Perturbation} variant leads to a 3.48\% AUC drop, indicating that static hidden states alone are insufficient for reliable request decisions. The \textit{w/o Sliding Window} variant decreases AUC by 2.06\%, suggesting that maintaining local spatial continuity around salient GUI regions is important. Naive strategies, such as \textit{Gaussian Noise} and \textit{Random Masking}, also degrade performance because they lack visual priors, while \textit{Inverse Saliency} performs comparably to the \textit{w/o Perturbation} variant. These results confirm that saliency-guided perturbation is critical for exposing latent cognitive instability.

\subsubsection{Impact of Trajectory Analysis.}
The \textit{w/o Trajectory Predictor} variant, which makes request decisions directly from hidden-state variance, causes a substantial AUC drop as it struggles to characterize complex latent drift. Replacing our predictor with mean-pooled or concatenated trajectories followed by an MLP reduces AUC by 4.99\% and 7.13\%, respectively. Mean pooling obscures divergence among trajectories, whereas concatenation lacks an explicit mechanism for modeling their relational structure. In contrast, the self-attention mechanism captures pairwise interactions among perturbed trajectories, enabling our predictor to quantify their latent dispersion more effectively.

\subsubsection{Impact of Trajectory Extraction.}
We further investigate how cognitive trajectory extraction affects the trajectory predictor by examining the extraction layer and token aggregation strategy, as detailed in Table~\ref{tab:extraction_strategy}.

We evaluate layer selection by extracting trajectories from different depths of the language model. The first and middle layers achieve AUCs of only 58.21\% and 66.85\%, respectively, indicating that shallow representations lack sufficient reasoning information to reflect actual cognitive confidence. The last layer also performs slightly worse than the penultimate layer with a 0.76\% AUC drop, likely because it is specialized for next-token prediction, while the penultimate layer preserves richer high-level semantic information. For token aggregation, the last-token embedding outperforms average pooling by 3.41\% in AUC. As the information summary before generation, the last token naturally aggregates critical cross-modal semantics from the input sequence, while average pooling mixes informative and background tokens, diluting these specific contextual semantics needed to capture cognitive instability.

\section{Conclusion}
We proposed SPECTRA, a lightweight autonomous request framework for edge-cloud GUI grounding. SPECTRA addresses the challenge where edge agents exhibit overconfident hallucinations in visually dense scenarios, causing a misalignment between confidence and accuracy that hinders reliable requesting. To this end, Saliency-Guided Targeted Perturbation constructs adversarial binary masks on the identified visual anchors to simulate stress tests on visual cognition. Efficient Cognitive Trajectory Analysis then quantifies the topological divergence of latent cognitive trajectories during the prefill phase. Extensive experiments across multiple benchmarks demonstrate that SPECTRA achieves superior request decision accuracy without autoregressive decoding. We believe that this paradigm of active cognitive stress testing offers a promising direction for reliable edge-cloud collaboration in GUI scenarios.

\begin{acks}
This work was supported by the National Natural Science Foundation of China (Grant Nos. 62402429, U24A20326, and 62441236), the Key Research and Development Program of Zhejiang Province (Grant No. 2025C01026), the Ningbo Yongjiang Talent Introduction Programme (Grant No. 2023A-397-G), and the Young Elite Scientists Sponsorship Program by CAST (Grant No. 2024QNRC001). The authors gratefully acknowledge support from the Qizhen Scholar Foundation of the Zhejiang University Education Foundation.
\end{acks}

\bibliographystyle{ACM-Reference-Format}
\bibliography{ref}

\end{document}